\documentclass[sigconf]{acmart}
\usepackage{bbold}
\usepackage{float} 
\usepackage{multirow}
\usepackage{subcaption}
\usepackage{balance}
\AtBeginDocument{
  \providecommand\BibTeX{{
    \normalfont B\kern-0.5em{\scshape i\kern-0.25em b}\kern-0.8em\TeX}}}

\copyrightyear{2023} 
\acmYear{2023} 
\setcopyright{acmlicensed}\acmConference[WWW '23 Companion]{Companion Proceedings of the ACM Web Conference 2023}{April 30-May 4, 2023}{Austin, TX, USA}
\acmBooktitle{Companion Proceedings of the ACM Web Conference 2023 (WWW '23 Companion), April 30-May 4, 2023, Austin, TX, USA}
\acmPrice{15.00}
\acmDOI{10.1145/3543873.3584626}
\acmISBN{978-1-4503-9419-2/23/04}

\begin{document}

\title[Bootstrapping Contrastive Learning Enhanced Music Cold-Start Matching]{Bootstrapping Contrastive Learning Enhanced Music Cold-Start Matching}

\author{Xinping Zhao}
\email{zhaoxinping@zju.edu.cn}
\affiliation{%
  \institution{Zhejiang University}
  \institution{NetEase Cloud Music, NetEase Inc.}
  \streetaddress{}
  \city{Hangzhou}
  \country{China}
}

\author{Ying Zhang$^*$}
\email{zhangying5@corp.netease.com}
\affiliation{%
  \institution{NetEase Cloud Music, NetEase Inc.}
  \streetaddress{}
  \city{Hangzhou}
  \country{China}
}

\author{Qiang Xiao}
\email{hzxiaoqiang@corp.netease.com}
\affiliation{%
  \institution{NetEase Cloud Music, NetEase Inc.}
  \streetaddress{}
  \city{Hangzhou}
  \country{China}
}

\author{Yuming Ren}
\email{renyuming@corp.netease.com}
\affiliation{%
  \institution{NetEase Cloud Music, NetEase Inc.}
  \streetaddress{}
  \city{Hangzhou}
  \country{China}
}

\author{Yingchun Yang$^*$}
\email{yyc@zju.edu.cn}
\affiliation{%
  \institution{Zhejiang University}
  \streetaddress{}
  \city{Hangzhou}
  \country{China}}

\thanks{$^*$Corresponding authors: Ying Zhang and Yingchun Yang.}

\renewcommand{\shortauthors}{Xinping Zhao et al.}

\begin{abstract}
We study a particular matching task we call Music Cold-Start Matching. In short, given a cold-start song request, we expect to retrieve songs with similar audiences and then fastly push the cold-start song to the audiences of the retrieved songs to warm up it. However, there are hardly any studies done on this task. Therefore, in this paper, we will formalize the problem of Music Cold-Start Matching detailedly and give a scheme. During the offline training, we attempt to learn high-quality song representations based on song content features. But, we find supervision signals typically follow power-law distribution causing skewed representation learning. To address this issue, we propose a novel contrastive learning paradigm named \textbf{B}ootstrapping \textbf{C}ontrastive \textbf{L}earning (BCL) to enhance the quality of learned representations by exerting contrastive regularization. During the online serving, to locate the target audiences more accurately, we propose \textbf{C}lustering-based \textbf{A}udience \textbf{T}argeting (CAT) that clusters audience representations to acquire a few cluster centroids and then locate the target audiences by measuring the relevance between the audience representations and the cluster centroids. Extensive experiments on the offline dataset and online system demonstrate the effectiveness and efficiency of our method. Currently, we have deployed it on NetEase Cloud Music, affecting millions of users. 
\end{abstract}



\maketitle
\section{Introduction \label{SEC1}}
Recommender systems (RS) play an important role in alleviating information overload and providing personalized services for large-scale online platforms \cite{covington2016deep, cheng2016wide, koren2009matrix, zhou2018deep}. However, in our real-world scenario, these elaborate models trained on enormous user-item interactions are powerless when facing cold-start songs with no past information\cite{pan2019warm}. We hope to warm up the cold-start songs quickly to solve this problem. And follow this line of thinking, we attempt to locate the target audiences of the cold-start song. But, there is a contradiction that the cold-start song hasn't any historical interactions so as hardly to discover the target audiences. Thus, we change our thoughts from a direct way to an indirect way. Given space limitations, we only introduce some key components. Specifically, we first aim to retrieve songs whose audiences have a high degree of overlapping with the cold-start song's audiences. Then, we fetch the audiences recently carrying out red-heart behavior \footnote{The red-heart behavior means clicking the Like button in NetEase Cloud Music.} on the retrieved songs as the candidate audience pool. Finally, we filter out the high-confidence target audiences from the candidate audience pool according to some methods and strategies. The working flow mentioned above is called Music Cold-Start Matching by us. 

For this purpose, we refer to the idea of content-to-collaborative filtering (CB2CF) \cite{barkan2019cb2cf} to develop XMusic \footnote{For copyright issues, we can't publish XMusic dataset and are sorry about that.}, a large-scale dataset composed of the song-to-song interactions. The build procedure of XMusic will be introduced in Section~\ref{SEC411}. Inspired by content-based music recommendation \cite{oramas2017deep, pulis2021siamese, chen2021learning, niyazov2021content}, we aim to leverage song content features, including audio, metadata, generated content, etc., to learn high-quality song representations supervised by the song-to-song interactions. However, we notice the imbalance of supervision signals between popular and less popular music, also known as the long tail effect. This phenomenon will cause insufficient representation learning for long-tail music \cite{wang2021contrastive, wei2022contrastive}. Inspired by the immense success of contrastive learning (CL) in a wide range of domains including computer vision (CV) \cite{chen2020simple, he2022masked, wang2021contrastive}, natural language processing (NLP) \cite{yan2021consert, zhou2022debiased, zhang2022contrastive}, recommender systems (RS) \cite{yao2021self, wu2021self, yu2022graph}, etc., we expect to introduce the CL’s superiority into Music Cold-Start Matching to learn more robust song representations. To fulfill the above goal, we propose BCL a novel contrastive learning paradigm consisting of two key mechanisms: (1) Correlation Grouping Mechanism (CGM), which divides features into different augmented groups; (2) Correlation Bootstrapping Mechanism (CBM), which dynamically updates the feature correlation matrix in a self-guided manner. Borrowing the idea of mask image modeling (MIM) in CV \cite{he2022masked}, we devise three feature-level data augmentation operators — random mask, span mask, and uniform noise — in order to perturb song representations in different manners. With model training finished, we run model inference to get the entire song pool and dynamically maintain it. Given a cold-start song request, we use the model to infer its representations and then efficiently retrieve top-k nearest songs from the pool by Faiss \cite{johnson2019billion}. Then, we query the database to fetch recent red-heart audiences of retrieved songs as the candidate audience pool. However, directly distributing the cold-start song to the candidate audiences is less elaborate and may influence the key indicators of the online system. To address this issue, we propose CAT a clustering-based method that first clusters audience representations into a few cluster centroids by K-means and then locates the target audiences by measuring the relevance between the audience representations and the cluster centroids. Finally, the cold-start song will be distributed to the target audiences to warm up it. The main contributions of this work are three folds:
\begin{itemize}
\item To the best of our knowledge, we are the first to formalize the problem of Music Cold-Start Matching and give a scheme. 
\item We propose BCL a contrastive learning paradigm to learn more robust song representations and CAT a clustering-based method to locate the target audiences more accurately.
\item The effectiveness and efficiency of our method are verified not only by the offline dataset but also by the online system, one of China’s biggest online music platforms\footnote{https://music.163.com/.}, contributing to significant business revenue growth.
\end{itemize}

\section{PRELIMINARIES}
We first formalize the workflow of the main supervised learning (SL) task. There are three categories of song content features \footnote{As for the image feature, we eliminate it because it only brings trivial influence.} in our scenario: attribute $a \in \mathcal{A}$, audio $o \in \mathcal{O}$, and lyric $l \in \mathcal{L}$. For attribute features, we employ the embedding layer to map the original high-dimensional attribute features into low-dimensional attribute representations $a = [a_1, a_2, ..., a_{n_a}] \in \mathbb{R}^{n_a \times d}$, where $n_a$ denotes the number of attribute features and $d$ denotes the embedding dimension. For the audio feature, we employ YAMNet \footnote{An audio event classifier: https://tfhub.dev/google/yamnet/1.} to extract the audio representation $o \in \mathbb{R}^d$. For the lyric feature, we encode it by the pre-trained language model BERT \cite{kenton2019bert} and take the transformed representation of the $[\rm{CLS}]$ token from the last layer as the lyric representation $l \in \mathbb{R}^d$. We fine-tune YAMNet and BERT during the training phase. Finally, the input can be represented as $x = [a, o, l] = [x^{(1)}, x^{(2)}, ... , x^{(k)}] \in \mathbb{R}^{k \times d}$, where $k = n_a+2$.

\begin{figure}[htbp]
	\centering
	\includegraphics[width=1.0\linewidth]{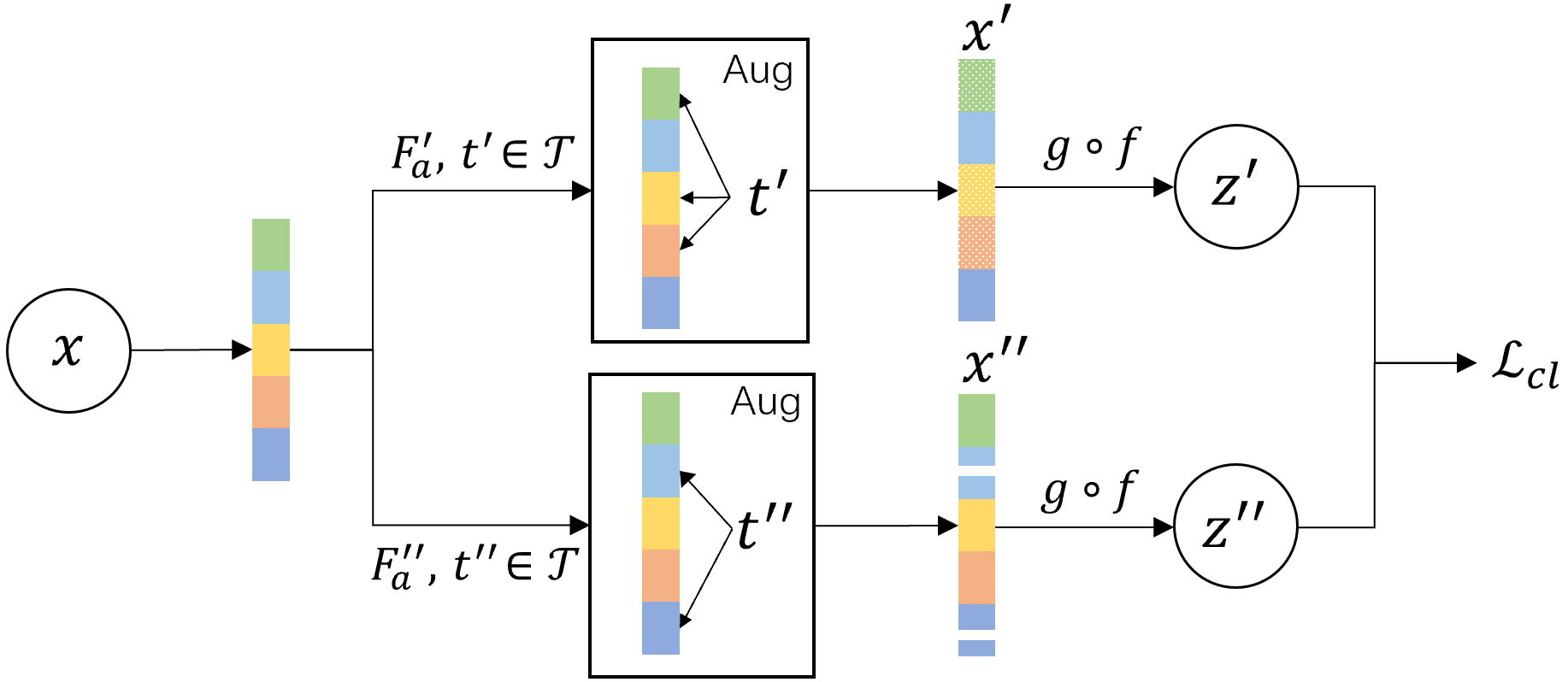}
	\caption{The overall working flow of BCL.}
    \Description[The overall working flow of BCL]{BCL first samples two augmentation operations and generates two augmented groups, applied to the input features $x$ to establish augmented views. Then, a backbone network encoder $f(\cdot)$ and a projection head $g(\cdot)$ are built upon the augmented views to extract representations. Finally, contrastive learning loss uses to refine the representations.}
	\label{fig1}
\end{figure}

A backbone network encoder $f(\cdot)$ is adopted to extract song representations $r \in \mathbb{R}^{d_r}$ based on the input $x$. Then, we employ the pairwise Bayesian Personalized Ranking (BPR) loss \cite{rendle2009bpr} to enforce the matching score of a coupled song-to-song interaction higher than its uncoupled counterpart:
\begin{equation}
  \mathcal{L}_{bpr} = \sum_{(i,j,k) \in \mathcal{D}} -\log(\sigma(r_i^T r_j - r_i^T r_k)),
\end{equation}
where $\mathcal{D} = \{ (i,j,k) | (i,j) \in \mathcal{D}^{+}, (i,k) \in \mathcal{D}^{-} \}$ is the training data, $\mathcal{D}^{+}$ is the coupled song-to-song interactions, and $\mathcal{D}^{-}$ is the sampled uncoupled song-to-song interactions.

\section{METHODOLOGY}
\subsection{Bootstrapping Contrastive Learning}
\subsubsection{Correlation Bootstrapping Mechanism}
We measure the feature correlation by distance correlation. In particular, distance correlation can measure both linear and nonlinear relationships of any two paired matrices, which can be formulated as:
\begin{equation}
  \mathcal{C}_{i,j} = dCor(X^{(i)}, X^{(j)}) = \frac{dCov(X^{(i)}, X^{(j)})}{\sqrt{dVar(X^{(i)})  \cdot  dVar(X^{(j)})}},
\end{equation}
where $X^{(i)}$ and $X^{(j)}$ denote the i-th and j-th feature matrices, $dCov(\cdot)$ is the distance covariance between two matrices, $dVar(\cdot)$ is the distance variance of each matrix. To adapt to the shift in feature distribution, we design a bootstrapping mechanism CBM enlightened by \cite{liu2022towards, caron2018deep}. The core idea of our solution is to dynamically update the feature correlation matrix $\mathcal{C}$ every $k$ steps via a slow-moving average amendment of the feature correlation matrix $\mathcal{S}$ calculated from the current input data: 
\begin{equation}
  \mathcal{C} = \alpha \mathcal{C} + (1-\alpha) \mathcal{S},
\end{equation}   
where $\alpha \in [0, 1]$ is a hyper-parameter to control the speed of the exponentially weighted moving average.

\begin{figure}[htbp]
	\centering
	\includegraphics[width=1.0\linewidth]{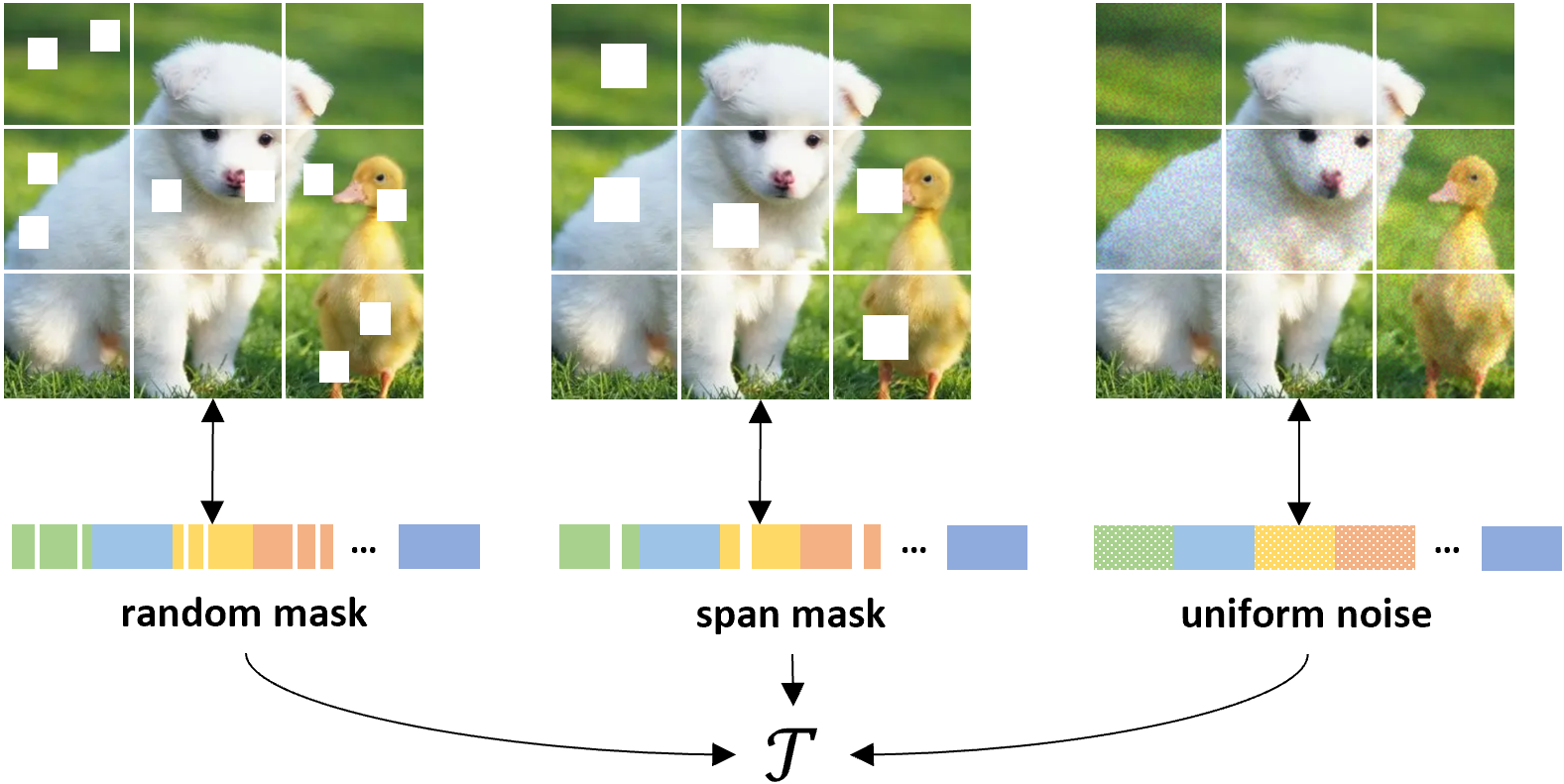}
	\caption{A vivid illustration of augmentation operations.}
     \Description[Three augmentation operations in our BCL]{In our BCL, there are three augmentation operations are used to establish augmented views, i.e., random mask, span mask, and uniform noise.}
	\label{fig2}
\end{figure}

\subsubsection{Correlation Grouping Mechanism} To seek more meaningful and challenging feature-dependency patterns, we design a grouping mechanism CGM enlightened by \cite{yao2021self} and beyond it. Specifically, we first uniformly sample a seed feature $f_{s}$ from all the candidate features $F = \{f_1, ..., f_k\}$. Then, we adopt the Gumbel-Max trick \cite{gumbel1954statistical, maddison2014sampling} to sample $n$ different features $F_{s} = \{ f_{s,1}, ..., f_{s,n}\}$ according to their correlation with the seed feature $f_{s}$: 
\begin{equation}
  \underset{i}{\arg\max}\big(\log C_{s,i} - \log(-\log \pi_i) \big)_{i=1,i \neq s}^{k},
\end{equation} 
where $\pi_i$ obey the uniform distribution $U(0, 1)$ and $n$ is equal to $ \lfloor \frac{k-1}{2} \rfloor$. Finally, we set the union of the seed feature $f_s$ and the $n$ sampled features $F_s$ as one augmented group $F_a^{\prime} = \{ f_{s}, f_{s,1}, ..., f_{s,n} \}$ and the rest as another augmented group $F_a^{\prime\prime} = \{f_{r,1}, ..., f_{r,(k-n-1)} \}$. It is noteworthy that we resample the seed feature $f_{s}$ for each training step to seek various meaningful and challenging feature-dependency patterns.

\subsubsection{Feature-level Data Augmentation} Inspired by MIM in CV \cite{he2022masked}, we devise three feature-level data augmentation operators as shown in Figure~\ref{fig2}: (1) Random Mask, masking each feature in the augmented group with randomly sampled positions; (2) Span Mask, masking each feature in the augmented group with continuous positions beginning from the sampled starting position; (3) Uniform Noise, adding imperceptibly small noises to each feature in the augmented group. Formally, we set $\rho$ to control the masking ratio and $\epsilon$ to control the magnitude of the noise.

\subsubsection{Contrastive Learning} A projection head $g(\cdot)$ is built upon the backbone encoder to map song representations $r$ to the suitable space where contrastive loss is applied. Then, we obtain a new vector representation $z \in \mathbb{R}^{d_z}$. To avoid skewed contrastive learning, we uniformly sample a minibatch of $N$ songs and establish $2N$ augmented views. After that, we treat the views of the same song as the positive pairs (i.e., $\{(z_i^{\prime}, z_i^{\prime\prime}) | 1 \leq i \leq N \}$), and the view of any different songs as the negative pairs (i.e., $\{(z_i^{\prime}, z_k^{\prime\prime}) | 1 \leq i,k \leq N, i \neq k \}$). Following SimCLR \cite{chen2020simple}, we adopt infoNCE \cite{gutmann2010noise} as the contrastive loss to maximize the agreement of positive pairs and minimize that of negative pairs:
\begin{equation}
  \mathcal{L}_{cl} = \sum_{i=1}^{N} - \log \frac{\rm{exp}(sim(z_i^{\prime}, z_i^{\prime\prime})/\tau)}{\sum_{k=1}^{N}  \mathbb{1}_{[k \neq i]} \rm{exp}(sim(z_i^{\prime}, z_k^{\prime\prime})/\tau)},
\end{equation}
where $\rm{sim}(\cdot)$ is the cosine similarity function, and $\tau$ is the temperature coefficient. The workflow of BCL is illustrated in Figure~\ref{fig1}.

\subsection{Multi-task training}
To enable the auxiliary CL task to help improve the learning of the main supervised task, we adopt a multi-task training strategy to jointly optimize these two objectives: 
\begin{equation}
  \mathcal{L} = \mathcal{L}_{bpr}+\lambda_1\mathcal{L}_{cl}+\lambda_2\| \Theta  \|_2^2,
    \end{equation}
where $\Theta$ denotes all trainable parameters, $\lambda_1$ and $\lambda_2$ are hyper-parameters to control the strengths of CL loss and $L_2$ regularization.

\subsection{Clustering-based Audience Targeting} To locate the target audiences more accurately, we propose CAT. Specifically, we first adopt the widely used K-means to conduct clustering among the candidate audience pool to find some representative centroids as the weak classifiers. Then, we refer to the bagging strategy, ensembles of multiple weak classifiers to reduce the variance of the results. Next, we measure the relevance scores between the candidate audience representation and the representative centroids by the cosine similarity function. It is noteworthy that the audience representations are fetched from the feature platform. For each candidate audience, we aggregate its relevance scores with each representative centroid in a weighted sum manner, where the weight is proportional to the magnitude of the cluster. Finally, we take the top-m most related audiences as the target audiences based on their weighted relevance score, where m is a pre-defined value.

\begin{table}
  \caption{Statistics of the dataset.}
  \Description[The detailed statistics of the XMusic dataset]{There are 318 thousand songs and 5.4 million song-to-song interactions in the XMusic dataset. The average interactions for each song are 17.26 and the density of the dataset is 0.005\%.}
  \label{tab1}
  \begin{tabular}{c|c|c|c|c}
    \toprule
    Dataset & \# songs & \# interactions & avg interactions & density\\
    \hline
    \texttt{XMusic} & 318K & 5.4M & 17.26 & 0.005\%\\
    \bottomrule
  \end{tabular}
\end{table}

\section{EXPERIMENTS}
\subsection{Experimental Settings}

\subsubsection{Dataset \label{SEC411}} We refer to the idea of CB2CF \cite{barkan2019cb2cf} to develop XMusic dataset. Specifically, we collect collaborative signals (billion-level) over 90 days from multiple key scenarios, e.g., DailySong and FM. Then, we analyze the song-to-song cooccurrence in the user behavior sequence, where the behavior not only denotes the play behavior but also contains other explicit behavior, e.g., the red-heart behavior and the songmark behavior. Next, we compute a cooccurrence score for each song-to-song pair in line with our purpose. Finally, we filter out high-confidence song-to-song pairs according to the pre-defined threshold. The detailed statistics of the dataset are summarized in Table~\ref{tab1}.

\subsubsection{Compared methods} To verify the effectiveness, we compare our BCL with the following non-CL and CL methods: (a) Base refers to the vanilla backbone network; (b) Feature Dropout (FD) \cite{volkovs2017dropoutnet} exerts random feature dropout on input to condition for missing preference patterns; (c) Global Orthogonal Regularization (GOR) \cite{zhang2017learning} impose global orthogonal regularization on input to maximize the "spread-out" property in the descriptor space; (d) Correlated Feature Masking (CFM) \cite{yao2021self} apply correlated feature masking on input to learn the better latent relationship of item features.

\subsubsection{ Implementation details and Metrics} For a fair comparison, we implement all models with the same architecture \footnote{The backbone network $f(\cdot)$ and the projection head $g(\cdot)$ are implemented by a three-layer MLP and a one-layer MLP equipped with one hidden layer to purely verify the effectiveness of the method.} initialized by the Xavier method \cite{glorot2010understanding} and we train them by Adam \cite{kingma2014adam} with learning rate of 0.001 and mini-batch size of 1024. The embedding size is 128, and the weight decay $\lambda_2$ is $1e^{-4}$. For other hyper-parameters, we employ the grid search to tune them and report the best result. We adopt Recall@50 and NDCG@50 as evaluation metrics and evaluate the ranking results over the entire song set.

\begin{table}[htbp]
  \caption{Overall results of different models trained on 10\% and full training data. The best results are boldfaced, and the second-best results are underlined. }
  \Description[Overall results of different models trained on 10\% and full training data]{Experimental results show our BCL achieves significant improvements whether under the condition of data sparsity or not. Compared with the state-of-the-art method CFM, BCL achieves improvements of 4.74\% and 8.74\% in terms of Recall@50 and NDCG@50, as per full training data.}
  \label{tab2}
  \begin{tabular}{c|c|c|c|c}
    \toprule
     & \multicolumn{2}{c|}{10\% XMusic Dataset} & \multicolumn{2}{c}{ Full XMusic Dataset} \\ \hline
      Method & Recall@50 & NDCG@50 & Recall@50 & NDCG@50 \\ \hline
      Base & 0.0477 & 0.0288 & 0.0619 & 0.0407 \\ 
      FD & 0.0478 & 0.0284 & 0.0622 & 0.0392 \\ 
      GOR & 0.0494 & 0.0292 & 0.0646 & 0.0411 \\ 
      CFM & \underline{0.0543} & \underline{0.0308} & \underline{0.0696} & \underline{0.0435} \\ 
      BCL & \textbf{0.0613} & \textbf{0.0355} & \textbf{0.0729} & \textbf{0.0473} \\ \hline \hline
      \%Improv. & 12.89\% & 15.26\% & 4.74\% & 8.74\% \\
    \bottomrule
  \end{tabular}
\end{table}

\begin{figure}[htbp]
	\centering
	\includegraphics[width=1.\linewidth]{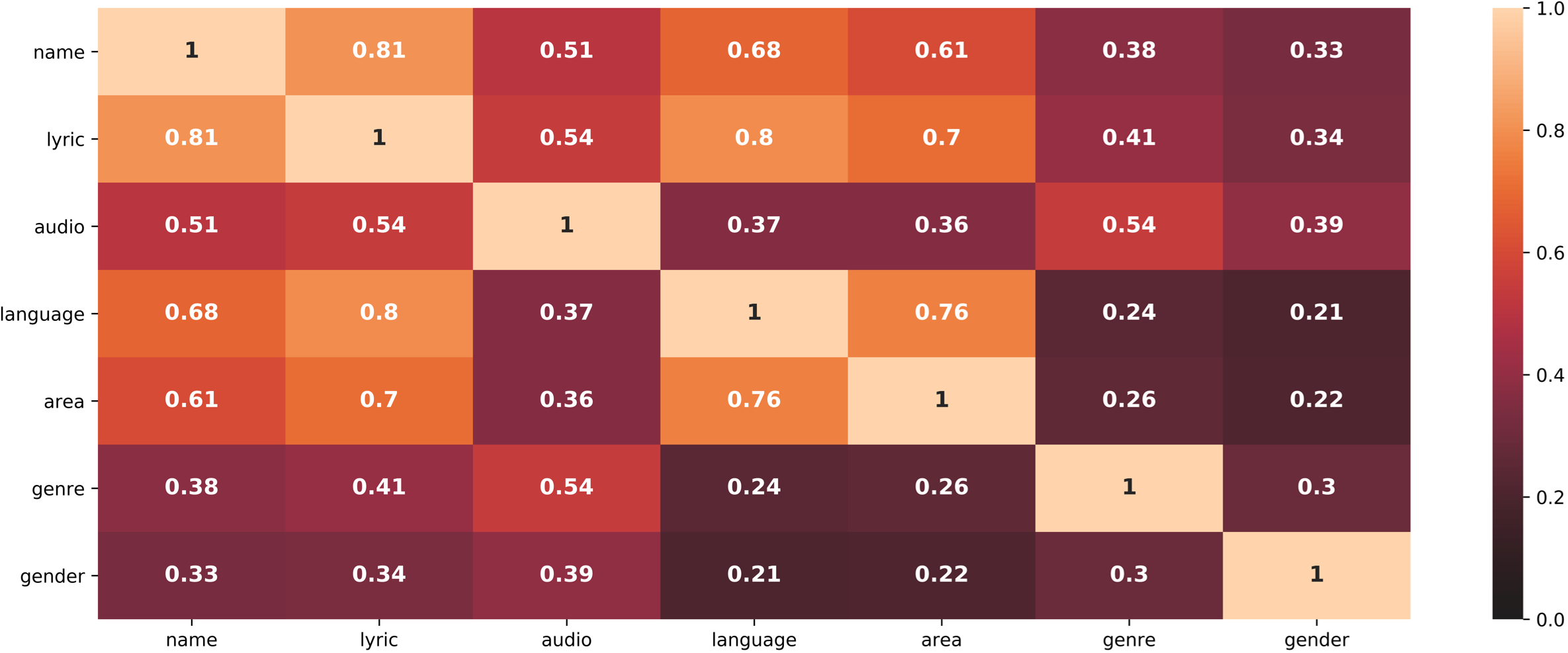} 
	\caption{ The heatmap of the feature correlation matrix learned by BCL. The higher the number, the stronger the correlation.}
    \Description[The heatmap of the feature correlation matrix learned by BCL]{The feature correlation matrix indeed captures accurate correlations conforming to human intuition. The audio feature, for example, its most correlated features are the lyric and genre features, which conform to the actual situation. }
	\label{fig3}
\end{figure}

\begin{figure}[htbp]
	\centering
	\includegraphics[width=1.0\linewidth]{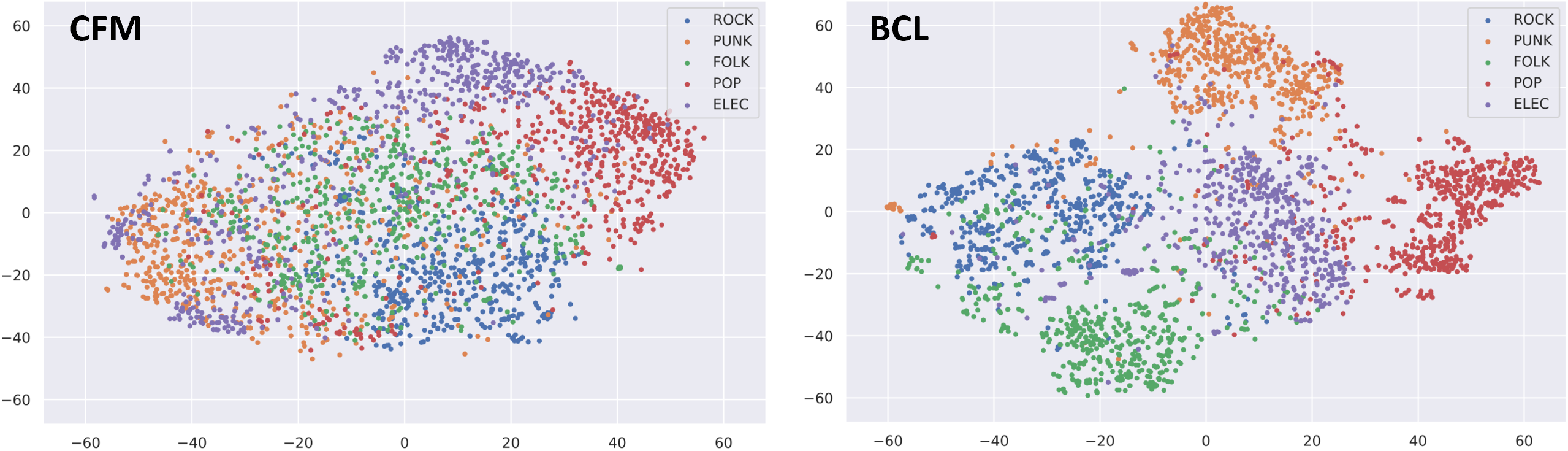} 
	\caption{ Comparison of t-SNE plots for song representations.}
    \Description[Comparison of t-SNE plots for song representations]{In the comparison of t-SNE plots for song representations learned by CFM and BCL, BCL learns more intra-genre compact and inter-genre separable song representations than CFM.}
	\label{fig4}
\end{figure}

\begin{table}[htbp]
  \caption{Online A/B test of BCL compared to Base.}
  \Description[Online A/B test of BCL compared to Base]{The introduction of BCL brings +38.47\% huge improvements in effective play count compared to Base. Meanwhile, BCL doesn’t degrade the user experience. On the contrary, it slightly boosts word-of-mouth, with +1.42\% improvements in full play rate and +1.95\% improvements in effective red rate.}
  \label{tab3}
  \begin{tabular}{c|c|c|c}
    \toprule
     & effective play count & full play rate & effective red rate \\ \hline
    Gain & +38.47\% & +1.42\% & +1.95\% \\ 
    \bottomrule
  \end{tabular}
\end{table}

\begin{figure}[htbp]
	\centering
	\includegraphics[width=1.0\linewidth]{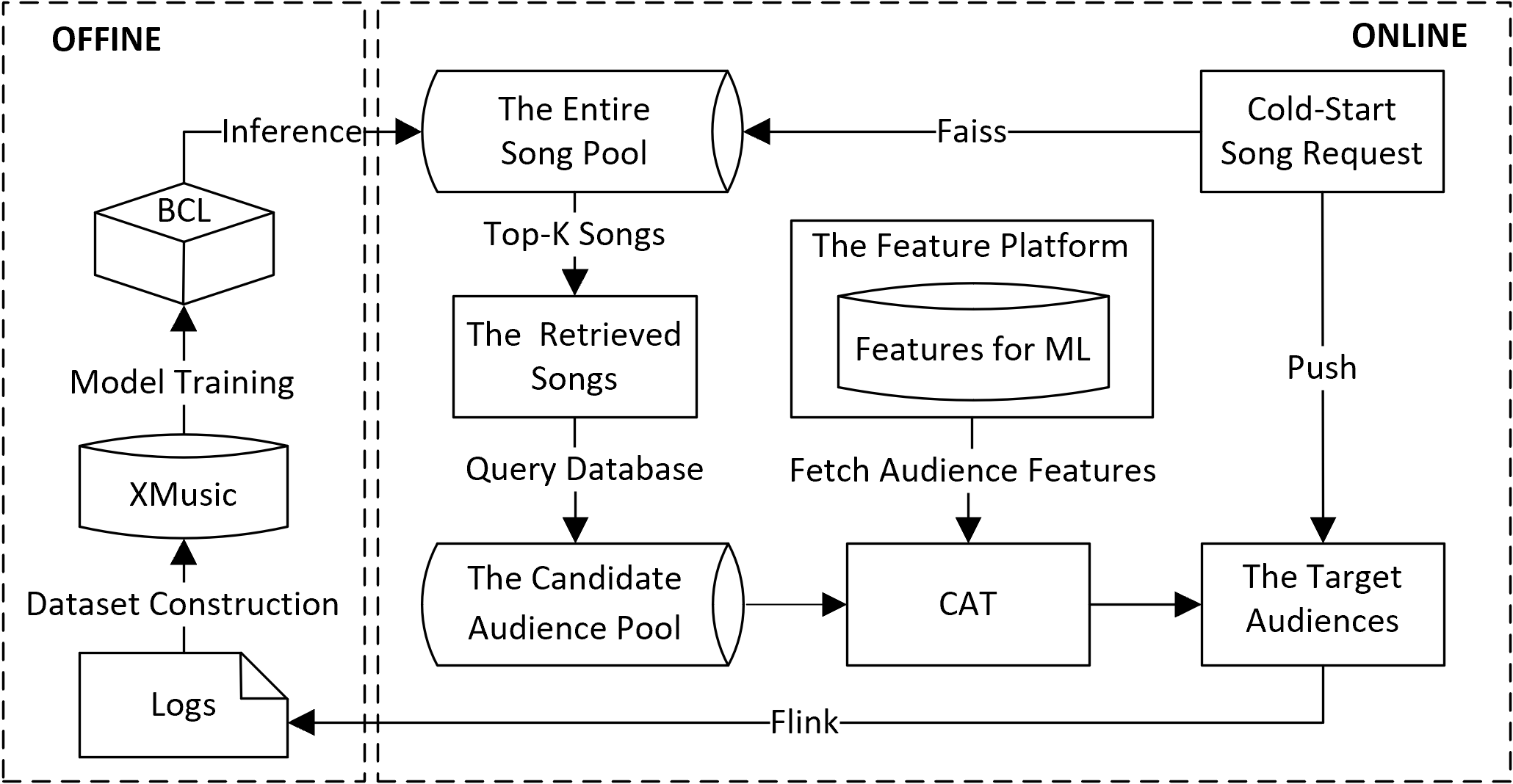} 
	\caption{The pipeline of Music Cold-Start Matching. The left is the offline stage while the right shows the online stage.}
     \Description[The pipeline of Music Cold-Start Matching.]{The pipeline of Music Cold-Start Matching contains offline and online stages. In the offline stage, the system attempts to learn high-quality song representations. In the online stage, the system aims to locate the target audiences for a given cold-start song request and distribute it to the target audiences to warm up it.}
	\label{fig5}
\end{figure}

\subsection{Offline Experiments}
To explore the impact of data sparsity on model performance, we train all models not only on full XMusic training data but also on 10\% down-sampled training data while all of them are evaluated on the full test data. As illustrated in Table~\ref{tab2}, our method achieves significant improvements, which fully validate its effectiveness. By further analyzing the experimental results, we have several observations: (1) For the methods without CL loss, including Base and FD, BCL outperforms them by a large margin, which indicates the superiority of assisting the SL task with CL; (2) For the methods with CL loss, including GOR and CFM, BCL consistently outperforms them by a considerable margin. Specifically, BCL achieves improvements of 4.74\% and 8.74\% in terms of Recall@50 and NDCG@50, compared to CFM, as per full training data. We think the main reason is BCL can dynamically mine the meaningful and challenging feature-dependency patterns among heterogeneous features favoring representation learning; (3) Under the setting of 10\% training data (highly sparse), BCL can also achieve a fairly good performance, which means BCL can mitigate data sparsity issues effectively. On the other hand, we choose seven typical features to facilitate the display and visualize their feature correlation matrix by heatmap as shown in Figure~\ref{fig3}. We can clearly see that the feature correlation matrix indeed captures accurate correlations conforming to human intuition. Furthermore, we visualize the song representations of five relatively popular music genres learned by CFM and BCL with t-SNE as shown in Figure~\ref{fig4}. We can see BCL learns more intra-genre compact and inter-genre separable song representations than CFM.

\subsection{Online A/B test}
To verify the effectiveness of BCL in real-world scenarios, we conduct an online A/B test on NetEase Cloud Music from July 4 to July 10 in 2022. In online evaluation, we focus on the following three indicators: (1) effective play count, (2) full play rate, (3) effective red rate, where the "effective" means play duration greater than a pre-defined threshold, the "full" means playing a song from beginning to end, and the "red" means clicking the Like button. As shown in Table~\ref{tab3}, the introduction of BCL brings +38.47\% huge improvements in effective play count compared to Base, which satisfies our goal of warming up the cold-start songs quickly. Meanwhile, BCL doesn't degrade the user experience. On the contrary, it slightly boosts word-of-mouth, with +1.42\% improvements in full play rate and +1.95\% improvements in effective red rate. Now, as shown in Figure~\ref{fig5}, we have deployed BCL to serve as the module of music push. The deployment pipeline has been detailed in Section~\ref{SEC1}.

\section{CONCLUSION}
In this work, we formalize the problem of Music Cold-Start Matching and give a scheme. To enhance the quality of the learned representations, we put forward a novel contrastive learning paradigm BCL to exploit various meaningful and challenging feature-dependency patterns in a self-guided manner. In addition, we design a clustering-based method CAT to locate the target audiences more accurately in an ensemble manner. Both offline and online experiments demonstrate the effectiveness and efficiency of our method. At present, we have deployed it on NetEase Cloud Music, contributing to significant business revenue growth.

\clearpage
\bibliographystyle{ACM-Reference-Format}
\balance
\bibliography{reference}

\end{document}